# Nonlinear optical Hall effect in Weyl semimetal WTe$_2$

Young-Gwan Choi[1][†], Manh-Ha Doan[1][†], Youngkuk Kim[2,3], and Gyung-Min Choi[1,3]*

[1] Department of Energy Science, Sungkyunkwan University, Suwon 16419, Korea

[2] Department of Physics, Sungkyunkwan University, Suwon 16419, Korea

[3] Center for Integrated Nanostructure Physics, Institute for Basic Science, Suwon 16419, Korea

[†]These authors contribute equally

*E-mail: gmchoi@skku.edu

**The ordinary Hall effect refers to generation of a transverse voltage upon exertion of an electric field in the presence of an out-of-plane magnetic field. While a linear Hall effect is commonly observed in systems with breaking time-reversal symmetry *via* an applied external magnetic field or their intrinsic magnetization[1, 2], a nonlinear Hall effect can generically occur in non-magnetic systems associated with a nonvanishing Berry curvature dipole[3]. Here we report, observations of a nonlinear optical Hall effect in a Weyl semimetal WTe$_2$ without an applied magnetic field at room temperature. We observe an optical Hall effect resulting in a polarization rotation of the reflected light, referred to as the nonlinear Kerr rotation. The nonlinear Kerr rotation linearly depends on the charge current and optical power, which manifests the fourth-order nonlinearity. We quantitatively determine the fourth-order susceptibility, which exhibits strong anisotropy depending on the directions of the charge current and the light polarization. Employing symmetry analysis of Berry curvature multipoles, we demonstrate that the**

**nonlinear Kerr rotations can arise from the Berry curvature hexapole allowed by the crystalline symmetries of WTe$_2$. There also exist marginal signals that are incompatible with the symmetries, which suggest a hidden phase associated with the nonlinear process.**

Whereas the ordinary Hall effect of electrons discovered in 1879 by Edwin H. Hall commonly arises in a two-dimensional conductor under a perpendicular magnetic field, the Hall effect has become a ubiquitous phenomenon, occurring in various systems ranging from magnetic insulators to non-magnetic semiconductors[1,2]. Notably, the nonlinear Hall effect has been theoretically predicted and experimentally observed in a *non-magnetic* system under a time-reversal symmetric condition[3-6]. This finding stands a contrast to the *a priori* sense that the Hall effect should occur without time-reversal symmetry, since time-reversal symmetry zeroes the Chern number, which dictates the quantized Hall conductance of the two-dimensional electrons[7, 8]. The key innovation was to reveal that it is only in the first-order response regime that time-reversal symmetry suppresses the Hall voltage, in which the Hall voltage is linearly proportional to a driving electric field (*E*-field), and the Hall conductivity is independent of *E*-field. In the case of a higher-order response regime, the presence of time-reversal symmetry rather allows for a new opportunity to encounter a novel Hall effect, namely a nonlinear Hall effect [3, 9-11]. Time-reversal symmetry in the absence of inversion symmetry enables a dipolar distribution of the Berry curvature, which leads to the nonzero integration of the Berry curvature gradient in momentum space, dubbed the Berry curvature dipole [3, 4]. Consequently, a second-order nonlinear Hall voltage dominates the Hall effect, induced by the Berry curvature dipole, and the Hall voltage (conductivity) features a quadratic (linear) dependence on *E*-field. Theoretically, it is possible that crystalline symmetries can further suppress the second-order nonlinear Hall effect with the help of time-reversal symmetry. Therefore, a higher-order nonlinear Hall effect can emerge[12-14], which is yet to be addressed experimentally.

The Hall effect is responsible for not only electrical responses but also optical responses of electrons, which is referred to as the optical Hall effect[15]. The linear optical Hall response is widely used to measure magnetization of a system *via* magneto-optic Faraday effect, magneto-optic Kerr effect, or magnetic circular dichroism[16]. When light passes through (Faraday rotation) or is reflected from (Kerr rotation) magnetized mediums, its polarization rotates. In this case, the rotation angle, which is linearly proportional to the Hall conductivity $\sigma_{xy}$, is independent on *E*-field of light, but depends on the light's frequency (*ω*). This polarization rotation arises owing to the optical Hall effect, in which an *E*-field of an incident light induces a transverse *E*-field. The optical detection of the nonlinear electrical Hall effect is also reported in a strained MoS$_2$ monolayer, where the Hall conductivity is driven by an external charge current and independent on *E*-field of light[17].

The optical Hall effect is an ideal platform to investigate high-order nonlinear processes owing to the intense *E*-field of a short-pulsed laser. A strong nonlinear optical effect has been often discovered with topologically non-trivial materials, associated with topological quantities, such as the Berry curvature, responsible for various nonlinear optical effects, including shift current, second-harmonic generation, and nonlinear Kerr rotation[12]. A two-dimensional massless Dirac system, such as graphene, has been extensively studied for exceptionally large nonlinear optical activities from terahertz to near infrared regime[18-20]. Recently, enhanced large nonlinear optical responses have been observed in three-dimensional topological Dirac/Weyl semimetals, e.g. photocurrent[21-23], second- or third-harmonic generation[12, 24], and optical Kerr effect[25, 26]. Furthermore, an anomalously large coefficient for the linear magneto-optic Kerr effect has been reported with a magnetic Weyl semimetal[27]. Tungsten ditelluride, WTe$_2$, known as type-II Weyl semimetal[28-30], can be a good platform to investigate the nonlinear optical response. Previously, the topological properties of WTe$_2$ have been studied mainly by electrical transport measurements, such as negative magnetoresistance, planar Hall effect, and electrical Hall effect [29, 31-34].

In this work, we demonstrate, for the first time, the current-induced nonlinear optical Hall effect in multilayer WTe$_2$ Weyl semimetal, which measures the analogue of the fourth-order nonlinear Hall effect under a time-reversal symmetric condition. The polarization rotation is measured at room temperature without application of magnetic field. In analogy with the magneto-optical Kerr rotation, which corresponds to a linear optical Hall effect, we call our observation as the nonlinear Kerr rotation. Our Kerr rotation should be distinguished from the well-known optical Kerr effect, which is 3$^{rd}$ order nonlinear effect but is not the Hall effect. The optical Kerr effect is driven by the diagonal part of the dielectric tensor, and its mechanism is related to the anharmonic potential of bound electrons. Whereas, our Kerr rotation is driven by the off-diagonal part of the dielectric tensor, and its mechanism is related to the Berry curvature (shown later). We give a precise fingerprint of the fourth-order nonlinearity by showing that the Kerr angle is proportional to the cube of the electric field: linear in the displacement field of charge current and quadratic in the proving field of the light. Our findings suggest that $WTe_2$ can be used for the electrically controllable nonlinear-optic-medium. The corresponding fifth-rank susceptibility tensor is quantitatively determined with different directions of the charge current and light polarization, leading to the maximum susceptibility of $\sim 10^{-24}$ m$^3$ V$^{-3}$. The fourth-order nonlinearity suggests the mechanism based on a non-vanishing Berry curvature hexapole. This predicts the dominant components of the susceptibility comprise those compatible with the crystalline symmetries of $WTe_2$, which is in line with the observations. Residual signals exist from symmetry-incompatible components, which can be an indicative of a possible hidden phase during the nonlinear process.

Our experiment is based on multilayer $WTe_2$ with thickness of 50~70 nm, exfoliated on $SiO_2$/Si substrates from $WTe_2$ bulk crystal (see Methods). Figure 1a shows a top view of its crystal structure. The crystal orientations ($a$- and $b$-axis directions) of the exfoliated flakes are precisely determined by polar Raman spectroscopy (Fig. 1b and Supplementary Information S1)[35, 36] and transmission electron microscope techniques (Fig. 1c and Supplementary

Information S2). The electrical electrodes are fabricated along the $a$- and $b$-axes of WTe$_2$ by electron-beam lithography following by a deposition of Cr/Au metals (Fig. 1d). Our fabricated devices show metallic behavior of WTe$_2$ with linear current-voltage characteristics (Fig. 1e) and the known temperature dependences of resistance (Fig. 1f). The resistance anisotropy ratio between the $b$- and $a$-axis directions is about 1.5, which is also in good agreement with the literature value[5, 37] (Supplementary Information S3).

For studying electro-optical phenomena in WTe$_2$ devices, we adopt the polarization rotation microscopy in a reflection geometry with an application of oscillating charge current $\vec{j}_c$ (Fig. 2a). A pulsed laser with a linear polarization, the center wavelength 785 nm ($\omega \sim$ 1.6 eV/$\hbar$), and the pulse duration ~100 fs is applied to the devices. After reflection from the devices, polarization variation, e.g. rotation and ellipticity angles, is monitored (see Methods). Current-induced Kerr rotation can be expressed as $\Delta\tilde{\theta}_K^\omega(\vec{j}_c) \equiv \tilde{\theta}_K^\omega(\vec{j}_c) - \tilde{\theta}_K^\omega(0)$, where $\tilde{\theta}_K^\omega(\vec{j}_e)$ and $\tilde{\theta}_K^\omega(0)$ are complex Kerr angles in the optical first-harmonic mode with and without the charge current, respectively (details in Supplementary Information S4). The $\vec{j}_c$ is converted to the displacement field $\vec{D}$ employing $\vec{D} = \rho \vec{j}_c$, where $\rho$ is the resistivity of WTe$_2$. (Note that $\vec{D}$ and $\vec{E}$ are used to differentiate the fields of the charge current and the incident light, respectively.)

We first demonstrate the fourth-order nonlinearity of the observed Kerr rotation. Figure 2b shows that the Kerr rotation is a linear function of $D_b$, which is the $b$-component of the displacement vector $\vec{D}$. Moreover, when the direction of the $D_b$ is reversed, the sign of Kerr angle direction is also changed, implying that the chirality of the system can be electrically switched. We note that the current-induced Kerr rotation only occurs with a pulsed laser but not with a continuous-wave laser of the same time-average power. This result suggests that the Kerr rotation is nonlinear in terms of the electric field of light $\vec{E}$. At the same time-average power, a femtosecond pulsed laser carries a much higher intensity of $\vec{E}$ compared to that of the continuous wave laser. Indeed, the Kerr rotation shows a linear dependence on the optical power of a pulsed laser. (The Kerr rotation is measured from a photocurrent of a balanced detector as $\Delta\tilde{\theta}_K^\omega(j_e) = \frac{\Delta I}{2I_0}$, where $I_0$ is the total photocurrent by reflected light and $\Delta I$ is the change of the photocurrent by the polarization rotation of light. The raw data of $\Delta I$ show a quadratic dependence on the optical power.) The linear dependence on the optical power indicates that $\Delta\tilde{\theta}_K^\omega(j_e) \propto E_i E_j$ as $I = nc\varepsilon_0|\vec{E}|^2$, where $n$ is the refractive index, $c$ is the speed of light, $\varepsilon_0$ is the vacuum permittivity. This result is in stark contrast to the linear optical Hall effect, dubbed magneto-optical Kerr rotation, in which the Kerr rotation is independent on $\vec{E}$. Based on these results, we conclude that the Kerr rotation of WTe₂ has a cube dependence on the fields $\Delta\tilde{\theta}_K^\omega(j_e) \sim D_b|\vec{E}|^2$, and thus demonstrates the fourth-order nonlinearity of the Hall effect.

Having established the fourth-order nonlinearity, we explore the anisotropic nature of the observed Kerr rotation. We perform a two-dimensional scanning of the Kerr rotation with different polarization directions of a linearly polarized light and displacement field as shown in Fig. 3. The Kerr rotation appears prominently throughout the sample area when the displacement field $\vec{D} \parallel \hat{b}$ and light field $\vec{E} \parallel \hat{a}$ (Fig. 3b). When both $\vec{D}$ and $\vec{E}$ are aligned along the $b$-axis ($\vec{D} \parallel \hat{b}$ and $\vec{E} \parallel \hat{b}$), there still exists finite uniform signal, but the intensity is significantly weakened compared to the first case (Fig. 3c). More importantly, when $\vec{D} \parallel \hat{a}$, it results in negligible signals, irrespective of the polarization direction as shown in Fig. 3e and Fig. 3f. Consequently, the Kerr rotation occurs prominently when $\tilde{\theta}_K^\omega(j_e) \sim D_b E_a^2$ and marginally when $\tilde{\theta}_K^\omega(j_e) \sim D_b E_b^2$. These features are stark contrast to the second-order nonlinear Hall effect, in which the displacement field in $\vec{D} \parallel \hat{a}$ gives a rise to the electric Hall

voltage.[4, 5]

We, then, quantitatively analyze the nonlinear Kerr rotation and make connection with the fourth-order nonlinear Hall conductivity. Under the action of the electric field $\vec{E}(\omega)$ of the light incident along the sample $\hat{c}$-direction, the Kerr rotation with an initial polarization along $a$- and $b$-directions can be expressed as,

$$\Delta \tilde{\theta}_K^{ba} \approx \frac{-2\varepsilon_{ba}}{(n_a + n_b)(1 - n_a)(1 + n_n)}, \tag{1}$$

$$\Delta \tilde{\theta}_K^{ab} \approx \frac{2\varepsilon_{ab}}{(n_a + n_b)(1 + n_a)(1 - n_b)}, \tag{2}$$

where $\varepsilon_{\mu\delta}$ ($\mu \neq \delta$), with subscript denotes the crystal axis, is the off-diagonal components of the dielectric tensor, which is related with the Hall conductivity as $\varepsilon_{\mu\delta} = \frac{1}{i\varepsilon_0 \omega}\sigma_{\mu\delta}$, and $n_\mu$ is the refractive index with the light polarization along the $\mu$-direction (see Supplementary Information S5 for the derivation.). Since the Kerr rotation comes from the fourth-order nonlinear process, we can express the $\varepsilon_{\mu\delta}$ in terms of a fourth-order susceptibility tensor $\chi^{(4)}_{\mu\alpha\beta\gamma\delta}$ as

$$\varepsilon_{\mu\delta} = \chi^{(4)}_{\mu\alpha\beta\gamma\delta} D_\alpha(0) E_\beta(\omega) E_\gamma^*(-\omega). \tag{3}$$

As an example, when $\vec{D}$ is applied along the sample $b$-axis, and $\vec{E}$ is along the $a$-axis (Fig. 3b), $\varepsilon_{ba} = \tilde{\chi}^{(4)}_{bbaaa} D_b E_a(\omega) E_a^*(-\omega)$ is responsible for the Kerr rotation. On the other hand, when all fields are aligned along the $b$-axis (Fig. 3c), the Kerr rotation arises from $\varepsilon_{ab} = \tilde{\chi}^{(4)}_{abbbb} D_b E_b(\omega) E_b^*(-\omega)$. Comparing the measured Kerr rotation (real and imaginary parts) with equations (1) and (2), we obtain $\tilde{\chi}^{(4)}_{bbaaa} = (6.2 - i1.6) \times 10^{-25}$ m$^3$ V$^{-3}$ and $\tilde{\chi}^{(4)}_{abbbb} = (-2.5 - i2.5) \times 10^{-25}$ m$^3$ V$^{-3}$. We also measure the Kerr rotation with arbitrary direction of

$E$ of light by changing the polarization direction in three-dimensional way: the azimuthal angle $\theta$ in the $ab$ plane; the polar angle $\psi$ in the $ab$ plane; the polar angle $\varphi$ in the $bc$ plane (Fig. 4). From the $\theta$ dependence of the Kerr rotation, we confirm the additional contribution from $\tilde{\chi}^{(4)}_{ababb}$ and $\tilde{\chi}^{(4)}_{bbbba}$. (To fit the experimental data, we use the Jones matrix formalism. See Supplementary Information S6). In addition, from the polar angle dependence, we identify contributions of $\tilde{\chi}^{(4)}_{bbaac}$ and $\tilde{\chi}^{(4)}_{abbbc}$ to the Kerr rotation. In particular, the magnitude of $\tilde{\chi}^{(4)}_{bbaac}$ of $> 2.7 \times 10^{-24}$ m$^3$ V$^{-3}$ is much larger than those of $\tilde{\chi}^{(4)}_{abbbb}$ and $\tilde{\chi}^{(4)}_{abbbb}$. As the fourth-order susceptibility is unexplored in other materials, to our best knowledge, it can be informative to convert the fourth-order susceptibility to the third-order susceptibility as $\tilde{\chi}^{(3)}_{baac} = \tilde{\chi}^{(4)}_{bbaac} D_b$. With $D_b$ of $6 \times 10^3$ V m$^{-1}$, we obtain $\tilde{\chi}^{(3)}_{baac}$ of $> 1.6 \times 10^{-20}$ m$^2$ V$^{-2}$, which is even larger than typical optical nonlinear systems, such as nanoparticle or polymer structures[38, 39]. Therefore, our result demonstrates a new functionality, electrical control of the optical nonlinear process, with topological materials.

Now, we provide a symmetry argument that explains the dominance of fourth-order nonlinear susceptibility $\chi^{(4)}_{\mu\alpha\beta\gamma\delta}$ for the Kerr rotation of WTe$_2$. Using the Boltzmann equation approaches, previous studies have shown that the fourth-order nonlinear susceptibility $\chi^{(4)}_{\mu\alpha\beta\gamma\delta}$ can be expressed as [13, 14],

$$\chi^{(4)}_{\mu\alpha\beta\gamma\delta}(\omega) = \frac{e^5}{i\varepsilon_0 \omega \hbar^4} \int \frac{d^3 k}{(2\pi)^3} f_{\text{FD}} \left( -i \frac{\partial_\alpha \partial_\beta \partial_\gamma \partial_\delta v_\mu}{\widetilde{3\omega}\,\widetilde{2\omega}\,\widetilde{\omega}^2} - \frac{\partial_\alpha \partial_\beta \partial_\gamma \Omega_{\mu\delta}}{\widetilde{2\omega}\,\widetilde{\omega}^2} \right), \tag{4}$$

where $e$ is the elementary charge, $\hbar$ is the reduced Planck constant, $f_{\text{FD}}$ is the Fermi-Dirac distribution, $\partial_\alpha \equiv \frac{\partial}{\partial k_\alpha}$, $\widetilde{n\omega} = n\omega + i/\tau$, $n$ is integer, $\omega$ is the angular frequency of light, $\tau$ is the relaxation time, and $\Omega_{\alpha\beta}$ is the Berry curvature with a pseudovector form of $B_\gamma = \frac{1}{2}\varepsilon_{\alpha\beta\gamma}\Omega_{\alpha\beta}$. The integration is done over the first Brillouin zone. The first term in the integrand is the Drude-like term, and the second term is related with the Berry curvature hexapole moment (see Supplementary Information S7 for the derivation). The existence of $\chi^{(4)}_{\mu\alpha\beta\gamma\delta}$ can be argued based on the symmetry consideration. The $T_d$-phase of WTe$_2$ belongs to an orthorhombic space group of P$mn2_1$ and point group of 2$mm$. It has a mirror reflection in the $bc$-plane ($M_a$) and a glide mirror reflection in $ac$-plane ($M_b$). It also preserves a time-reversal symmetry ($\mathcal{T}$) without an applied magnetic field, and the Drude-like term vanishes under the $\mathcal{T}$ symmetry. The Berry curvature hexapole gives four non-zero terms at the presence of the $M_a$, $M_b$, and $\mathcal{T}$ symmetry: $\partial_a\partial_a\partial_a\Omega_{ca}$, $\partial_a\partial_b\partial_b\Omega_{ca}$, $\partial_a\partial_a\partial_b\Omega_{cb}$, and $\partial_b\partial_b\partial_b\Omega_{cb}$. Specifically, the $M_a$ and $M_b$ symmetries force $\partial_\alpha\partial_\beta\partial_\gamma\Omega_{\mu\delta}$ to vanish when it contains an odd number of momentum index of $a$ or $b$. Since we measure the in-plane rotation ($\Delta\theta_K^{ba}$ or $\Delta\theta_K^{ab}$), only $\partial_a\partial_b\partial_b\Omega_{ca}$ and $\partial_a\partial_a\partial_b\Omega_{cb}$ terms are relevant. Then, the corresponding symmetry-allowed susceptibilities are $\tilde{\chi}^{(4)}_{abbca}$ and $\tilde{\chi}^{(4)}_{aabcb}$. If we allow the permutation among the subscript indexes, this prediction corresponds to $\tilde{\chi}^{(4)}_{bbaac}$, which is the largest susceptibility of our measurement. However, the symmetry consideration cannot explain other residual measurements, such as $\tilde{\chi}^{(4)}_{bbaaa}$ and $\tilde{\chi}^{(4)}_{abbbb}$. Note that even without the $M_b$ symmetry, which can be broken in a few layers of WTe$_2$ by the surface effect, $\tilde{\chi}^{(4)}_{bbaaa}$ and $\tilde{\chi}^{(4)}_{abbbb}$ should be vanished by the $M_a$ symmetry. As a possible explanation, we expect that a strong $E$-field of light might influence the crystal structure such that it spontaneously lowers the crystalline symmetry of WTe$_2$. For example, a back-gate bias was employed to induce the inversion symmetry breaking, which controls the Berry curvature dipole, in a monolayer and bilayer WTe$_2$[6, 23]. In addition, it has been reported that a terahertz light pulse can induce an ultrafast switch in symmetry of WTe$_2$ *via* lattice strain[40]. Furthermore, the semiclassical picture can potentially miss the possible quantum effects, notably associated with the multiband nature of WTe$_2$ at a sizable angular frequency $\hbar\omega = 1.6$ eV, as discussed by Parker *et. al* (ref. [14]). A detailed mechanism for the light-induced symmetry change would be an interesting topic for further study.

For a mechanism other than the Berry curvature multipoles, the chirality of the Weyl point in Weyl semimetals can directly induce nonlinear optical processes, such as photocurrent or circular photogalvanic effect[21-23]. Interestingly, ref. [23] reported a photocurrent response that is forbidden by the symmetry of a material of TaIrTe$_4$. Authors of ref. [23] interpreted the symmetry-forbidden response as a result of combination of the built-in electric field and optical excitation. Nonetheless, the chiral selection occurs with an optical transition from the lower part of the Weyl cone to the upper part, so it is mainly observed low photon energy. For connecting this effect to our work, further study is needed.

In summary, we observe the current-induced nonlinear optical Hall effect in WTe$_2$ multilayers at room temperature. The optical Hall effect is revealed as the nonlinear Kerr rotation, and the fourth-order nonlinearity is confirmed by the linear dependence on the charge current and optical power. Our work shows that the nonlinear optical process can be controlled by a small electric bias, thereby, demonstrates a useful functionality for the electro-optic device employing topological materials. From the anisotropic behavior of the Hall response, we quantitatively determine various the fourth-order susceptibilities ($\chi^{(4)}_{\mu\alpha\beta\gamma\delta}$). For the mechanism of $\chi^{(4)}_{\mu\alpha\beta\gamma\delta}$, we considered the Berry curvature hexapole, and it can explain the existence of $\tilde{\chi}^{(4)}_{bbaac}$, the largest susceptibility of our measurements. However, we found that some of our measurements, such as $\tilde{\chi}^{(4)}_{bbaaa}$ and $\tilde{\chi}^{(4)}_{abbbb}$, were forbidden by the symmetry of WTe$_2$. The observation of the symmetry forbidden terms suggests that not only a stable phase but also other unknown phases participate during the nonlinear process. We expect that the fourth-order nonlinear optical Hall effect could be a useful tool to investigate the topological nature of Weyl semimetal.

**Methods**

*Device Fabrication*: WTe$_2$ multilayer flakes are exfoliated from the WTe$_2$ bulk crystal onto SiO$_2$/Si substrates by using the scotch tape method inside a glove box with an argon environment to prevent oxidation. The sample is then coated by a PMMA layer before being taken out for Raman investigation and device fabrication. The in-plane crystal orientation (a- and *b*-axis) of the WTe$_2$ flake is determined by using an angle dependent polarized Raman spectroscopy (Supplementary Information S1). Cr/Au metal electrodes are fabricated along these two axes by the e-beam lithography and e-beam evaporation processes, respectively. The device is then etched by a reactive ion etching (RIE) system using SF6 gas at high vacuum to form a well-defined shape and is immediately loaded into the vacuum chamber of a RF sputtering system for coating a thin SiN passivation layer to prevent surface degradation of WTe$_2$ during the electrical and optical measurements. The thickness of the investigated devices is from 50 to 70 nm, which is confirmed by the atomic force microscopy (AFM). After all electrical and optical investigations, the crystal orientation of the fabricated devices is reconfirmed by the transmission electron microscopy (TEM) measurements (Supplementary Information S2).

*Electrical transport measurement:* The electrical characteristics of the fabricated device are initially analyzed at room temperature under a high vacuum (~ 10$^{-6}$ Torr) in a probe station system with a Keithley-4200SCS parameter analyzer. The device is then bonded to a puck and measured in a physical property measurement system (PPMS, Quantum Design) to investigate the temperature-dependent transports. The resistance of the sample could be measured at different temperatures which could be controlled at static or continuous scanning values in the ranges of 2-300 K (Supplementary Information S3).

*Current-induced Kerr rotation microscopy:* A linearly polarized pulsed laser with the center wavelength of 785 nm and a pulse duration ~100 fs is used as the excitation source. After reflection of the linearly polarized light from the sample, the polarization variation, e.g. the rotation and the ellipticity angles, can be monitored. For the normal incidence geometry, the light polarization angle $\theta$ is defined with respect to the *a*-axis. The reflected laser beam from the sample is passing through a half-wave plate, Wollaston prism, and the balanced detector for monitoring polarization variation with/without quarter-wave plate before it. Polarization variation can be precisely detected with a high sensitivity by using alternative bias induced modulation and a lock-in amplifier for increasing signal-to-noise ratio. The bias-induced Kerr angle can be expressed as, $\Delta\tilde{\theta}_K = \tilde{\theta}_K(j_c) - \tilde{\theta}_K(0)$, where $\tilde{\theta}_K$ is the complex Kerr angle, $j_c$ is the electrical current density, and $\Delta\tilde{\theta}_K$ is the variation of complex Kerr angle with an application of electrical bias (for details, see Supplementary Information S4).

**Acknowledgements**

G.-M. C. acknowledges the National Research Foundation of Korea (NRF) grant funded by the Korea government (2019R1C1C1009199, 2018M3D1A1058793). M.-H. D. acknowledges the NRF (2019R1I1A1A01063236). Y. K. acknowledges the NRF (2020R1A4A307970711). This study was supported by the Institute for Basic Science of Korea (IBS-R011-D1). All authors thank to Jungcheol Kim and Hyeonsik Cheong in Sogang university for their help with polarized Raman spectroscopy. G.-M. C. thanks to Kyung-Jin Lee in KAIST for theoretical discussion.

**Author contributions**

G.-M. C supervises the project. Y.-G. C and M.-H. D design the experiment, conduct device fabrication and characterization, and data analysis. Y. K. provides a theoretical consideration of the Berry curvature multipoles. All authors discuss the results and write the manuscript.

**Competing interests**

The authors declare no competing interests

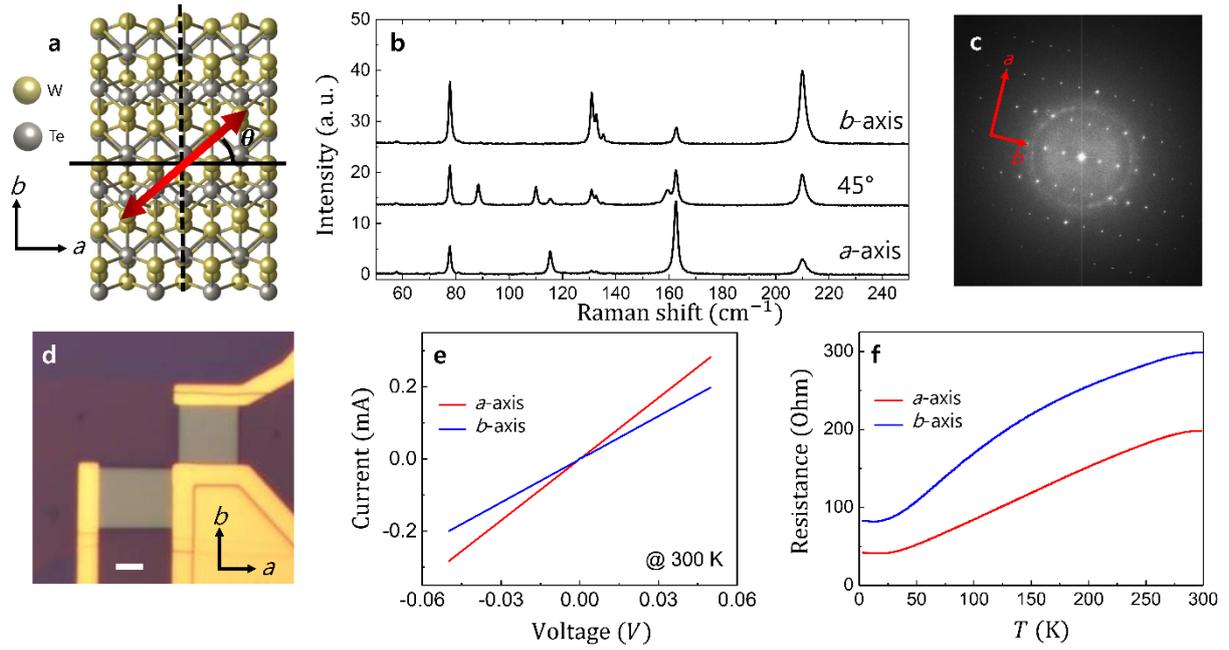

**Fig. 1| Crytal structure and metallic charateristic of WTe$_2$ multilayer. a,** Top-view of WTe$_2$ thin film. Red arrow indicates light polarization direction with an angle $\theta$ with respect to *a*-axis (solid line). Dashed line implies mirror plane. **b,** Polarization dependent Raman spectra along the different crystal directions of WTe$_2$. **c,** Fourier transform pattern from a transmission electron microscopy image of WTe$_2$ crystal showing the a- and b-axis directions. **d,** Optical microscope image of the investigated WTe$_2$ device. Scale bar is 5 μm. **e,** Current-voltage characteristics of the a- and *b*-axis devices at room temperature. **f,** Temperature-dependent resistivities of the devices showing the metallic behavior of WTe$_2$ multilayer.

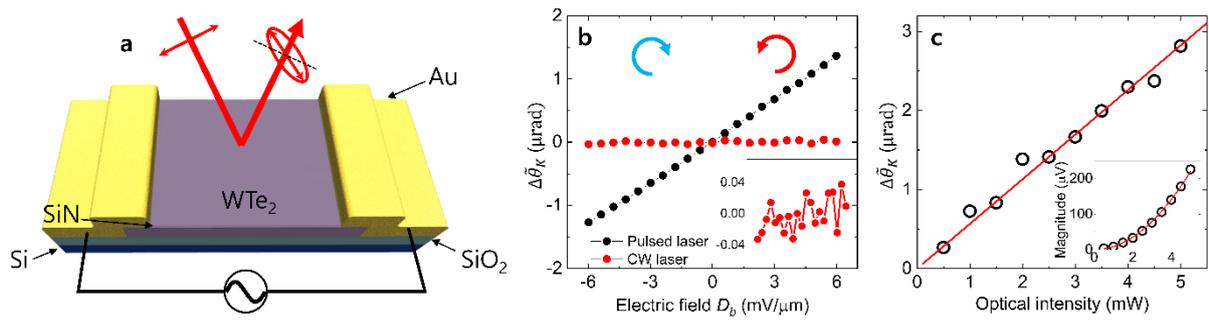

**Fig. 2| Current-induced nonlinear Kerr rotation in WTe$_2$ multilayer. a**, Schematic illustration of opto-electronic Kerr measurement for WTe$_2$ device. Red arrows indicate laser probing beam and its polarization directions. **b**, Kerr rotation as a function of the displacement field $D_b$. Kerr rotation is linearly proportional to $D_b$, and its sign is reversed when the direction of $D_b$ is opposite. The observed Kerr angle is much lower with continuous wave laser excitation (red open circles) compared to the case of pulsed laser excitation (black dots). The inset shows the enlarged graph of the continuous wave laser case. Optical power is the same for each case. **c**, Kerr rotation as a function of the optical power. The linear dependence on the optical power implies the optically nonlinear process. The inset shows that the raw signal ($\Delta I$) depends quadratically on the optical power.

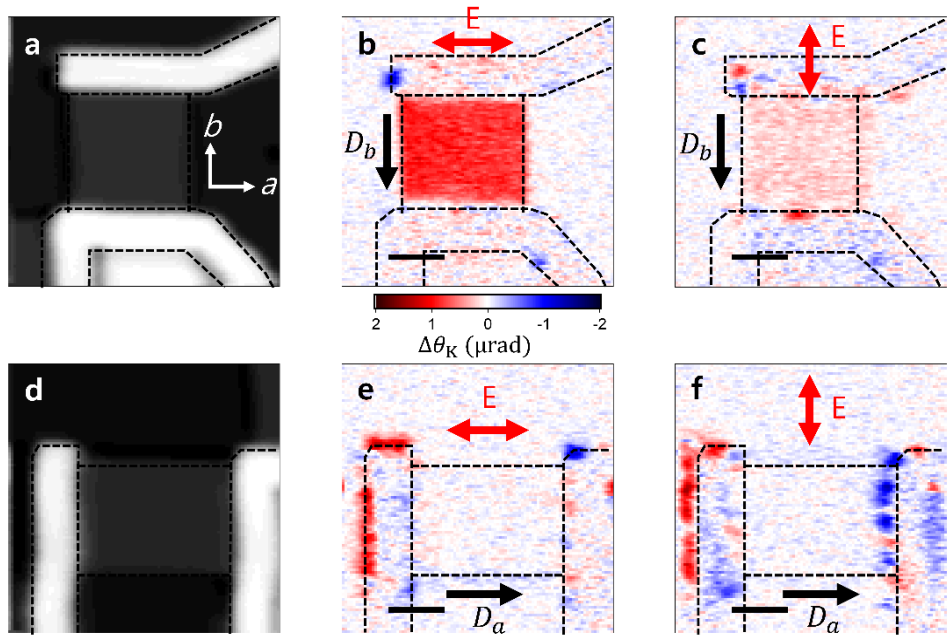

**Fig. 3| Spatial scan of Kerr rotation with different directions of charge current and light polarization. a,d**, Reflectance mapping images of the *b*-axis and *a*-axis devices. Scale bars indicate 5 μm. **b,c**, Kerr rotation mappings of the *b*-axis device with different directional light polarization **e,f**, Kerr rotation mappings of the *a*-axis device with the different directions of light polarization. Black and red arrows indicate the directions of the $\vec{D}$ of charge current and $\vec{E}$ of input light.

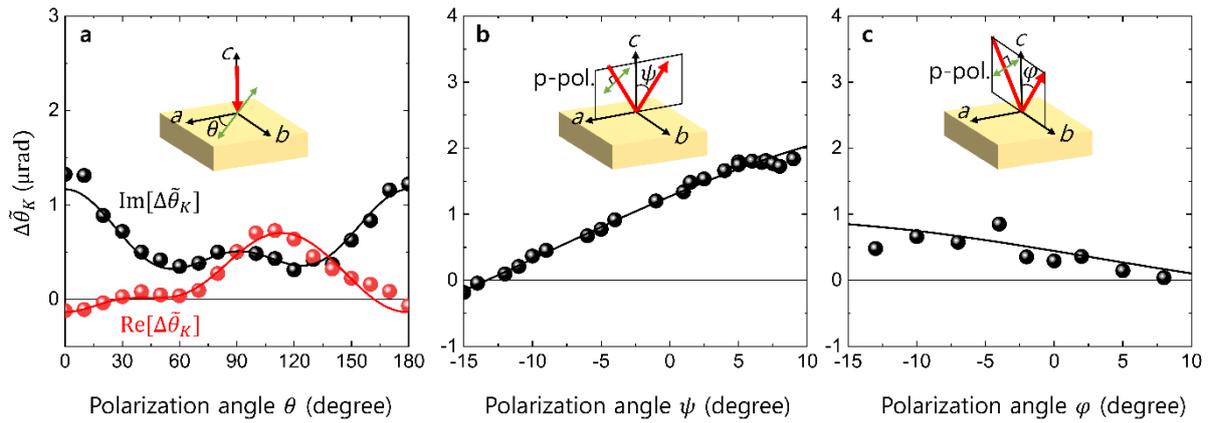

**Fig. 4| Kerr rotation with a three-dimensional variation of polarization of light. a**, Kerr rotation angle results as a function of azimuth angle, $\theta$, at a normal incidence geometry. Black and red data are imaginary and real part of the Kerr angle, respectively. **b,c**, Imaginary Kerr angle results as a function of grazing incidence angle $\psi$ and $\varphi$ at an oblique incidence plane of a-c and b-c, respectively, with p-polarization of incident light.